\documentclass{aipproc}
\usepackage{amsfonts}
\usepackage{amssymb}
\usepackage{epsf}

\input{aipcheck.tex}
\layoutstyle{6x9}

\begin{document}

\title{Four nucleon systems: a zoom to the open problems in nuclear
interaction} 
\author{Rimantas Lazauskas}{
  address={DPTA/Service de Physique Nucl\'eaire, CEA/DAM Ile de France, BP 12,
           F-91680 Bruy\`eres-le-Ch\^atel, France},
  email={lazauskas@lpsc.in2p3.fr}}

\iftrue
\author{Jaume Carbonell}{
  address={Laboratoire de Physique Subatomique et de Cosmologie, 53
  av. des Martyrs, 38026 Grenoble, France},
  email={carbonell@lpsc.in2p3.fr},
}
 \fi

\begin{abstract}
Faddeev-Yakubovski equations in configuration space are used to
solve four nucleon problem for bound and scattering states.
Different realistic interaction models are tested, elucidating
open problems in nuclear interaction description. On one hand, by
example of nonlocal Doleschall potential, we reveal possibility of
reducing three-nucleon force. On the other hand we disclose
discrepancies in describing n+$^3$H resonance, which seems to be
hardly related with off-shell structure of nucleon-nucleon
interaction.
\end{abstract}

\date{\today }
\maketitle

\copyrightyear  {2004}


Three and four nucleon systems are the cornerstones in
understanding nuclear interaction. The new generation of
nucleon-nucleon (\textit{NN}) potentials provide almost perfect
description of two-nucleon data, however most of them fail already
to describe the binding energies of \textit{3N} nuclei. It is
considered that off-shell effects, not comprised in local
\textit{NN} interaction models, are the cause of the underbinding
problem in the lightest nuclei. A standard practice to account for
the missing off-shell physics in description of \textit{N}>2
nuclei is implementation of \textit{3N} forces. In such a manner
it is not difficult to improve description of \textit{3N} and
\textit{4N} binding energies. However such procedure, which relies
only on good description of binding energies can be quite
misleading.

Low energy few nucleon physics is dominated by large
nucleon-nucleon scattering lengths and therefore, as can be
understood from effective range physics \cite{Platter}, the
binding energies of light nuclei depends only on very few
parameters. Thus study of nuclear binding energies, even being of
fundamental importance, can provide only very limited information
on nuclear interaction structure. Only at larger energies physical
observables should become sensible to particular form of the
potential. However at larger energies relativistic effects become
also stronger, while their impact is not very well understood. In
this scope \textit{4N} continuum, containing sensible structures
as thresholds and resonances presents a perfect ground to
understand nuclear interaction structure, see Fig. \ref{B_4N}.

Recently several theoretical techniques have enabled to solve
 \textit{4N} bound state problem accurately \cite{alpha_bench}.
 \textit{4N} continuum is more challenging task and advances in it are more modest.
 Still a few different groups and using different methods have been able to
obtain converged results for n+t elastic scattering with realistic
interactions  \cite{Viviani_cont,Bench,Hofmann}. Employed
formalisms enable to include Coulomb as well as to test different
\textit{NN} interactions, comprising nonlocal ones and models in
conjunction with three nucleon force (\textit{3NF}).
\begin{figure}\label{B_4N}
  \resizebox{18pc}{!}{\includegraphics{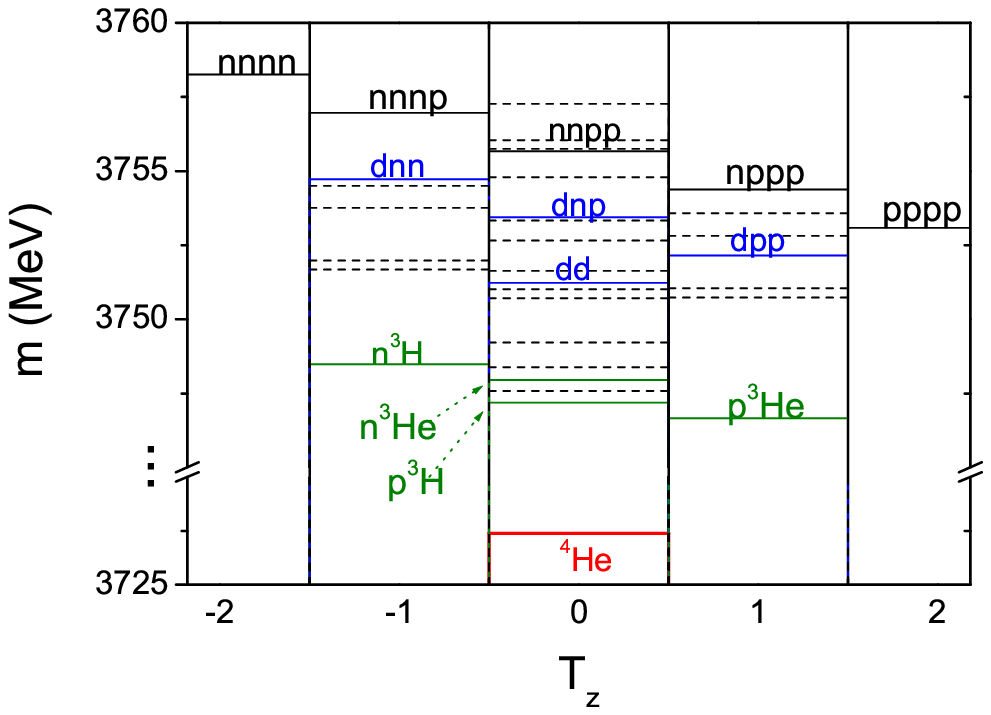}}
  \resizebox{18pc}{!}{\includegraphics{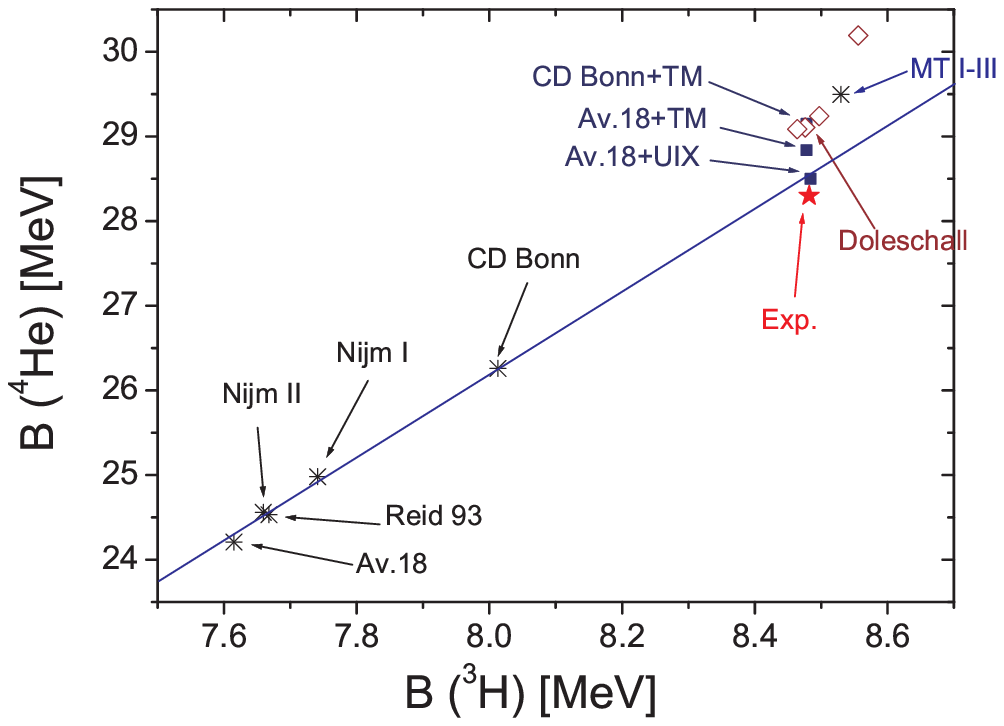}}
\caption{In left figure experimental spectra of \textit{4N} bound
and resonant states is presented. Figure on the right
recapitulates various predictions for $\alpha$-particle binding
energies (Tjon-line). Some energies in this figure are taken from
\cite{Nogga_alp}.}
\end{figure}

\subsection{Theoretical model}

To solve four-particle problem we use the Faddeev-Yakubovski
(\textit{FY}) equations in configuration space. In order to
include \textit{3NF  FY} equations are rewritten in form,
suggested by \cite{Gloeck_3BF}. For four identical particles,
these equations read:
\begin{eqnarray}
\left( E-H_{0}-V_{12}\right) K_{12,3}^{4} &=&V_{12}(P^{+}+P^{-})\left[
(1+Q)K_{12,3}^{4}+H_{12}^{34}\right] +V_{12,3}\Psi \\
\left( E-H_{0}-V_{12}\right) H_{12}^{34} &=&V_{12}\tilde{P}\left[
(1+Q)K_{12,3}^{4}+H_{12}^{34}\right]  \label{FY1}
\end{eqnarray}%
with $P^{+}$, $P^{-}$, $\tilde{P}$ and $Q$ being particle permutation
operators:
\begin{eqnarray}
\begin{tabular}{lll}
$P^{+}=(P^{-})^{-}=P_{23}P_{12};$ & $Q=\varepsilon P_{34}$; & $\tilde{P}%
=P_{13}P_{24}=P_{24}P_{13}$.%
\end{tabular}%
\end{eqnarray}%
and
\begin{eqnarray}
\Psi =\left[ 1+(1+P^{+}+P^{-})Q\right]
(1+P^{+}+P^{-})K_{12,3}^{4}+(1+P^{+}+P^{-})(1+\tilde{P})H_{12}^{34}
\end{eqnarray}%
the total wave function.

Equation (\ref{FY1}) becomes however non appropriate once long
range interaction, in particular Coulomb, is present. In fact,
\textit{FY} components remain coupled even in far asymptotes, thus
making implementation of correct boundary conditions hardly
possible. The way to circumvent this problem is in detail
described in \cite{Thesis}.

Equations (\ref{MFY_eq})in conjunction with the appropriate
boundary conditions are solved by making partial wave
decomposition of amplitudes $K_{12,3}^{4}$ and $H_{12}^{34}$:

\begin{eqnarray}
K_{i}(\vec{x}_{i},\vec{y}_{i},\vec{z}_{i}) &=&\sum_{LST}\frac{\mathcal{K}%
_{i}^{LST}(x_{i},y_{i},z_{i})}{x_{i}y_{i}z_{i}}\left[ L(\hat{x}_{i},\hat{y}%
_{i},\hat{z}_{i})\otimes S_{i}\otimes T_{i}\right] \\
H_{i}(\vec{x}_{i},\vec{y}_{i},\vec{z}_{i}) &=&\sum_{LST}\frac{\mathcal{H}%
_{i}^{LST}(x_{i},y_{i},z_{i})}{x_{i}y_{i}z_{i}}\left[ L(\hat{x}_{i},\hat{y}%
_{i},\hat{z}_{i})\otimes S_{i}\otimes T_{i}\right]
\end{eqnarray}

The partial components $\mathcal{K}_{i}^{LST}$ and $\mathcal{H}_{i}^{LST}$%
are expanded in the basis of three-dimensional splines. One thus
converts integro-differential equations into a system of linear
equations. More detailed discussion can be found in \cite{Thesis}.


\subsection{$\alpha$-particle}

The most general criticism of interaction models is based on their
ability to describe nuclear binding energies. Local \textit{NN}
interaction models require attractive contribution of \textit{3NF}
to shield $\sim$0.8 $MeV$ (or $\sim$10\%) underbinding in triton
and $^3$He binding energies. Usually, once these \textit{3N}
binding energies are fixed, one obtains rather good description of
the $\alpha$-particle. This is due to well known correlation
between $\alpha$-particle and \textit{3N} binding energies -- Tjon
line (see Fig. \ref{B_4N}) -- being a consequence of effective
range theory \cite{Platter}.  However the use of \textit{3NF}, to
some extent, can be just a matter of taste. In
fact, two different, but         
phase-equivalent, two-body interactions are related by an unitary,
nonlocal, transformation \cite{Polyzou_PRC58_98}. One thus could
expect that a substantial part of 3- and multi-nucleon forces
could also be absorbed by nonlocal terms. A considerable
simplification would result if the bulk of experimental data could
be described by only using two-body nonlocal interaction.

The inclusion of nonlocal terms in CD-Bonn model considerably
improves 3- and 4-\textit{N} binding energies, nevertheless this
improvement is still not sufficient to reproduce the experimental
values. A very promising result, which takes profit from
nonlocality in non relativistic nuclear models, has been obtained
by Doleschall and collaborators \cite{Dol_and_ref}. They have
managed to construct purely phenomenological nonlocal \textit{NN}
forces (called INOY), which are able to overcome the lack of
binding energy in three-nucleon systems, without explicitly using
\textit{3NF} and still reproducing \textit{2N} observables.
\begin{table}[h!]
\begin{tabular}{lcccc}
\hline
   \tablehead{1}{l}{b}{Pot. Model}
  & \tablehead{1}{c}{b}{$<T>$}
  & \tablehead{1}{c}{b}{$-<V>$}
  & \tablehead{1}{c}{b}{$B$}
  & \tablehead{1}{c}{b}{$R$ } \\
\hline
INOY96   & 72.45 & 102.7 & 30.19  & 1.358   \\
INOY03   & 69.54 & 98.79 & 29.24  & 1.373   \\
INOY04   & 69.14 & 98.62 & 29.11  & 1.377   \\
INOY04'  & 69.11 & 98.19 & 29.09  & 1.376   \\
AV18     & 97.77 & 122.1 & 24.22  & 1.516   \\
AV18+UIX & 113.2 & 141.7 & 28.50 \cite{Nogga_alp}  &   1.44
\cite{PPWC_PRC64_01}  \\\hline
Exp      &       &       & 28.30  & 1.47 \\
\hline
\end{tabular}
\caption{Binding energy B (in MeV) and rms radius R (in fm) for
$^4$He ground state obtained with Doleschall, AV18 and AV18+UIX
models.}\label{B_4N}
\end{table}

Using Faddeev-Yakubovski equations and by fully including Coulomb
interaction we have tested ability of INOY interaction models to
reproduce experimental $\alpha$-particle binding energy. In Table
\ref{B_4N} we summarize obtained results. One should mention that
the convergence of numerical results guarantied at least
three-digit accuracy. In contrary to \textit{3N} system,
experimental value of $\alpha$-particle binding energy is
overestimated by $\sim$800 keV. Nevertheless this discrepancy is
considerably smaller than for conventional local interaction
models and is on a par with the result obtained using CD-Bonn
\textit{NN} in conjunction with Tucson-Melbourne \textit{3NF}.

Still it is risky to judge about the interaction models basing
only on binding energy arguments, comparing other physical
observables can give us hints about the origin of the existing
drawbacks. In Table \ref{B_4N} one can clearly see that proton
r.m.s. radius in $\alpha$-particle predicted by INOY models is  by
$\sim$10\% smaller than the experimental one. Note, these models
provide experimental values of proton r.m.s in deuteron; in
3\textit{N} compounds these radii become slightly smaller than the
experimental ones \cite{CarbLaz_PRC04}. Such tendency demonstrates
that INOY interactions are too soft, resulting into a faster
condensation of the nuclear matter. In order to improve these
models one should try to increase the short range repulsion
between the nucleons.

This study reveals enormous liberty one has in describing
off-shell effects. Only careful analysis of continuum spectra,
which enables one to test various space configurations, can impose
stronger constrains to the off-shell structure of the nuclear
interaction.

\subsection{p+$^{3}$H scattering at very low energies
\label{Sec_4N_pt}}


$^{4}$He continuum is the most complex \textit{4N} system, its
spectrum contains numerous resonances (see Fig. \ref{B_4N}). In
this study we have considered only very low energy region
E$_{cm}$<500 keV region, below n+t threshold, fully described by
proton scattering in S-waves. Nevertheless excitation function -- $\left.{\frac{%
d\sigma}{d\Omega}}(E)\right|_{\theta=120^{\circ}}$ -- has
complicated structure due to existence of the $\mathcal{J}^{\pi
}=0^{+}$ resonance in p+$^3$H threshold vicinity.  This resonance,
being the first excitation of $\alpha$-particle, is located at
$E_{R}\approx0.4$ $MeV$ above p+$^{3}$H threshold and with its
width $\Gamma \approx0.5$ $MeV$ covers almost the entire region
below n+$^{3}$He.

\begin{figure}\label{nt_pt}
  \resizebox{20pc}{!}{\includegraphics{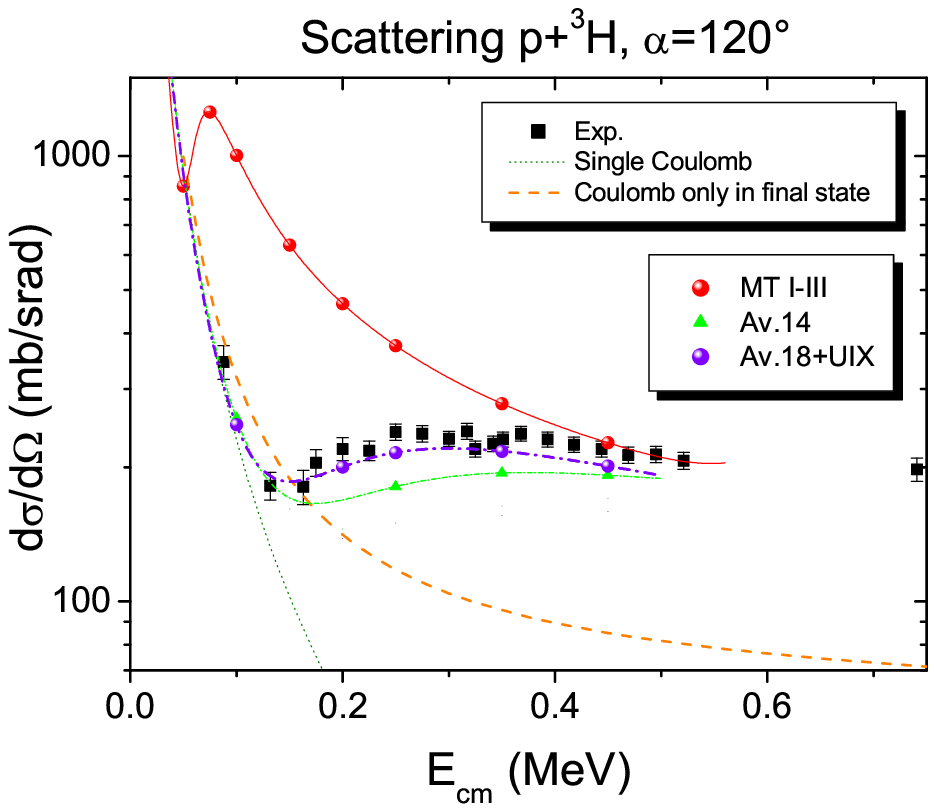}}
  \resizebox{20pc}{!}{\includegraphics{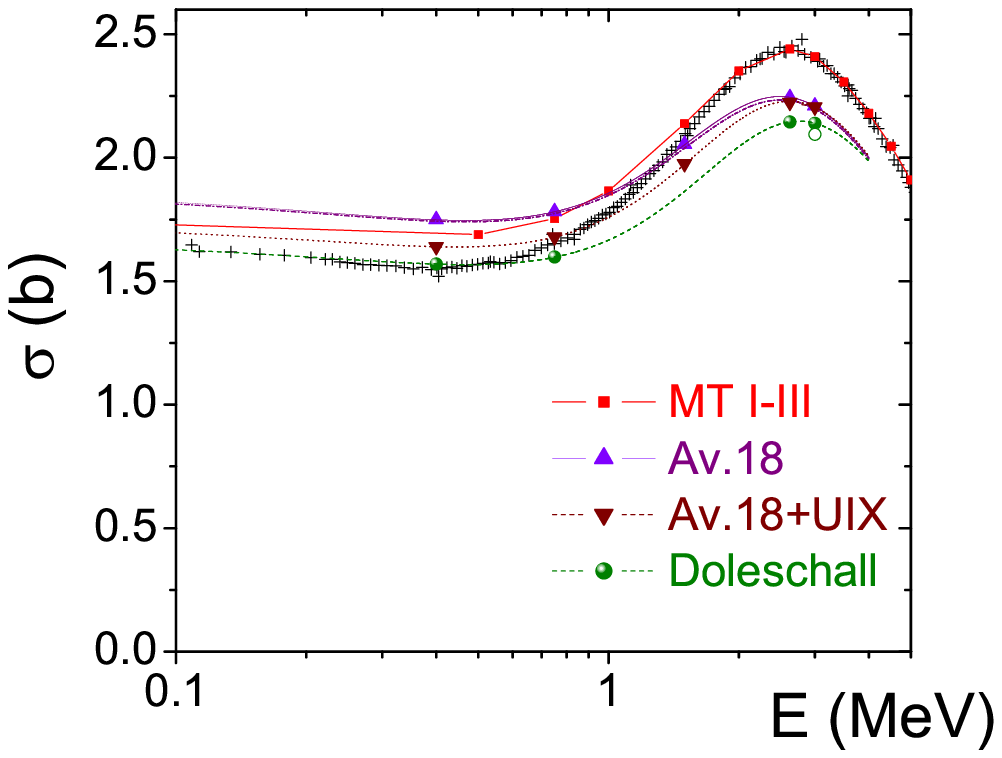}}
\caption{Various model calculations for p+$^{3}$H excitation function $\left.{\frac{%
d\sigma}{d\Omega}}(E)\right|_{\theta=120^{\circ}}$ (figure in the
left) and n+$^{3}$H cross sections (figure in the right).
Calculations compared to experimental data from \cite{pt_exp} and
\cite{nt_exp} respectively.}
\end{figure}

Separation of n+$^{3}$He and p+$^{3}$H channels requires proper
treatment of Coulomb interaction, the task is furthermore burden
since both thresholds are described by the same isospin quantum
numbers. When ignoring Coulomb interaction, as was a case in the
large number of nuclear scattering calculations, n+$^{3}$He and
p+$^{3}$H thresholds coincide. In this case $0^+$ resonant state
moves below the joint threshold and becomes bound. Former fact is
reflected in low energy scattering observables (see Fig.
\ref{nt_pt} dashed line): on one hand \textit{N}+\textit{NNN}
scattering length in $0^+$ state is found positive, on the other
hand excitation function decreases smoothly with incident
particles energy and does not show any resonant behavior. Only by
properly taking Coulomb interaction into account, thus separating
n+$^{3}$He and p+$^{3}$H thresholds, the $^4$He excited state is
placed in between and thus gives negative p+$^{3}$H singlet
scattering length.

In order to reproduce the shape of experimental excitation
function, \textit{NN} interaction model is obliged with high
accurately situate $^4$He excited state. In fact, width of the
resonance is strongly correlated with its relative position to
p+$^{3}$H threshold. If this state is slightly overbound the
resonance peak in excitation curve becomes too narrow and is
situated at lower energies. This is a case for MT I-III model
prediction, see Fig. \ref{nt_pt}. In case of underbinding one
obtains excitation function, which is too flat and provides
underestimated cross sections. This is a case of local
\textit{2NF} models, which predict too small (in absolute value)
singlet scattering length and thus place the virtual state too far
from threshold. Only once implementing UIX \textit{3NF} in
conjunction with Av18 \textit{NN} model one obtains singlet
scattering length as well as the excitation function  in agreement
with experimental data \cite{pt_exp}. On contrary various model
predictions do not differ much for triplet scattering case, where
resonance states are present only at considerably larger energies.

As have been shown, description of low energy p+$^{3}$H completely
relies on success in reproducing position and width of the first
excited state of $\alpha$-particle. However position of this
resonant state seems to be strongly correlated with position of
predicted triton threshold and $\alpha$-particle ground state. MT
I-III model overbinds a ground state of $\alpha$-particle, see
Fig. \ref{B_4N}, and thus its resonance is situated too close to
p+$^{3}$H threshold. On contrary local realistic interaction
models (without \textit{3NF}) strongly underbind $\alpha$-particle
and thus predict resonant state at higher relative energies to
triton threshold. Once triton and $\alpha$-particle energies are
reproduced (Av18+UIX case) one obtains also accurate description
for $^{4}$He excited state. These observations recalls to
effective range theory \cite{Platter}: in first order only one
three-body scale is important in describing three and four body
binding energies if two body physics is dominated by large
scattering lengths.

\begin{table}[tbh]\caption{\textit{4N} scattering lengths calculated using different
interaction models.} \label{table_a}

\begin{tabular}{lcccccccc}\hline
 \tablehead{1}{l}{b}{}&
 \tablehead{2}{c}{b}{MT I-III} & \tablehead{2}{c}{b}{Av14} &
\tablehead{2}{c}{b}{Av18+UIX}& \tablehead{2}{c}{b}{INOY04}\\

\textbf{System}& J$^{\pi }=0^{+}$ & J$^{\pi}=1^{+}$ & J$^{\pi
}=0^{+}$ & J$^{\pi }=1^{+}$ & J$^{\pi }=0^{+}$ & J$^{\pi }=1^{+}$&
J$^{\pi}=0^{+}$& J$^{\pi }=1^{+}$ \\\hline

\textbf{n-$^{3}$H} & 4.10 & 3.63 & 4.28 & 3.81 & 4.04 & 3.60&4.00  & 3.52\\
\textbf{p-$^{3}$H} & -63.1 & 5.50 & -13.9 &5.77 & -16.5 & 5.39& & \\
\hline

\end{tabular}
\end{table}

\subsection{n+$^{3}$H elastic scattering
\label{Sec_4N_nt}}

n+$^{3}$H elastic channel represents the simplest \textit{4N}
reaction. It is almost pure $\mathcal{T}=1$ isospin state, free of
Coulomb interaction in the final state as well as in the target
nucleus. Nevertheless this system contains two, spin degenerated,
narrow resonances at low energy ($E_{cm} \approx 3$ MeV), while
ability of nuclear interaction models to describe scattering cross
sections in this resonance region was put in doubt
\cite{Fonseca,Carbonell1}.
\begin{figure}\label{Fig_He_Exp_V18}
  \resizebox{18pc}{!}{\includegraphics{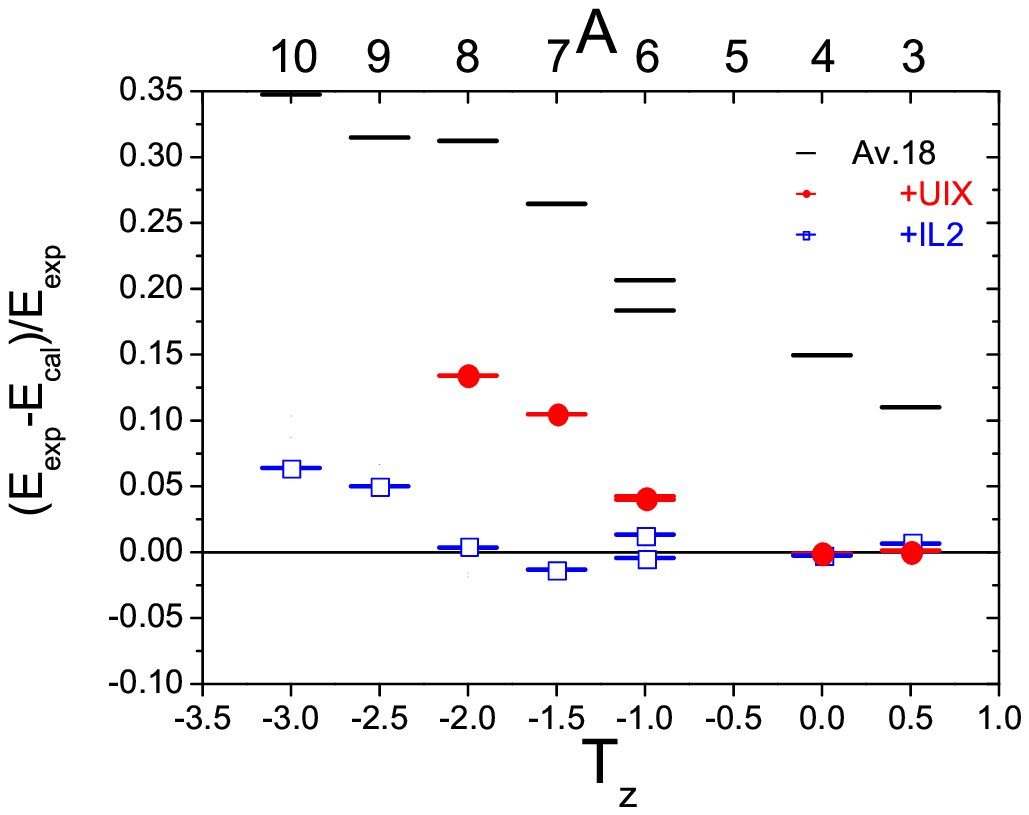}}
  \resizebox{18pc}{!}{\includegraphics{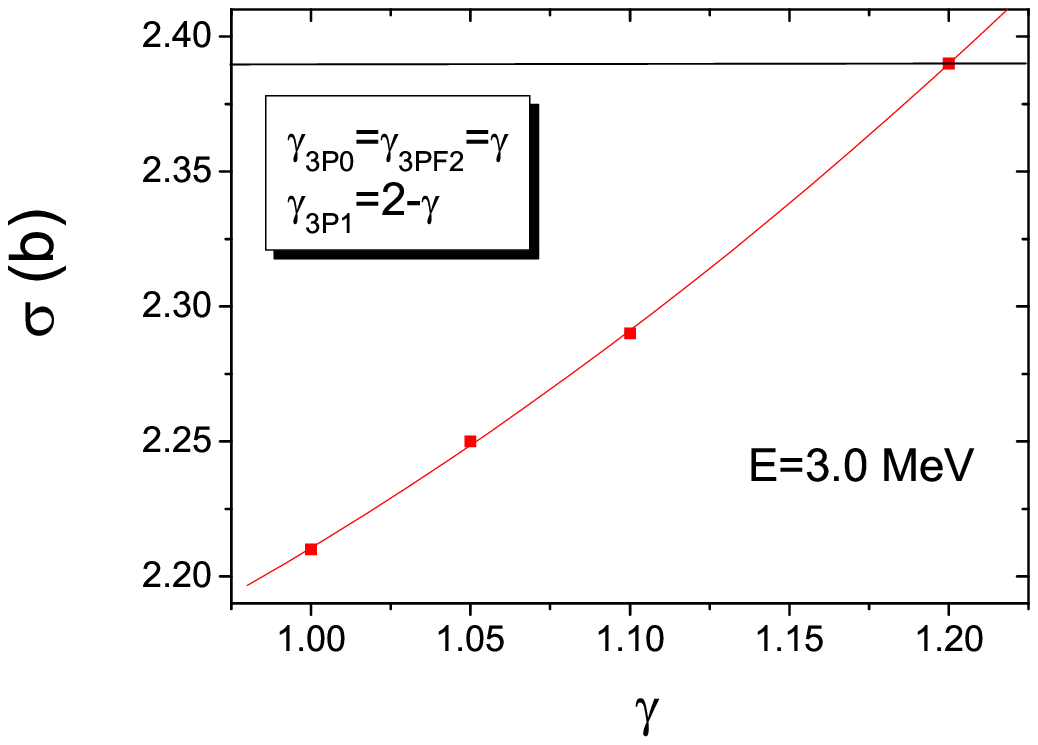}}
\caption{Comparison of calculated Ref.\cite{PPWC_PRC64_01} and experimental binding energies for various He isotopes 
(figure in the left). Right figure shows n+$^{3}$H elastic cross section at $E_{cm}=3$ MeV dependence on enhancement 
factor $\gamma$ of n-n P-waves.}
\end{figure}

Recently a few different groups have obtained the converged
results for n+$^{3}$H elastic scattering when using realistic
interaction models \cite{Viviani_cont}. Our predictions are in
full agreement with those of Pisa group, using Hyperspherical
Harmonics method, determining n+t scattering observables
 with accuracy better  than 1\% \cite{Bench}.

As have been shown before \cite{Viviani_PRL,Fonseca}, pure
\textit{NN} local interaction models overestimate n+$^{3}$H zero
energy cross sections, being consequence of overestimated size of
triton (lower binding energy). Scattering cross sections in this
very low energy region seems to be linearly correlated with triton
binding energy. However even implementing UIX \textit{3NF} for
AV18 model, and thus reproducing triton binding energy, small
discrepancies remain \cite{CarbLaz_PRC04}. If zero energy cross
section seems to fit experimental data, agreement for the coherent
scattering lengths is less obvious. Nonlocal interaction model
(INOY04) reproduces low n+$^{3}$H energy data better, this fact is
best reflected near the elastic cross section minima around
E$_{cm}\sim$0.4 MeV, see Fig. \ref{nt_pt}.

Nevertheless all realistic interaction models, including INOY,
underestimate by more than 10\% elastic cross sections in the
resonance region. Off-shell effects as 3NF or nonlocality of the
force, which improve low energy behavior, do not give sizeable
effect.

In order to understand the origin of this failure it is useful to
study impact of different potential terms for scattering
phaseshifts. This can be done using integral representation of the
phaseshifts. As was shown in \cite{CarbLaz_PRC04} for positive
parity states -- whether it is contribution in 3- and 4-\textit{N}
potential energies or in integral representation of scattering
phaseshifts -- \textit{NN} S-waves plays a major role, whereas
P-waves stay intact (less than 1\%). On the other hand
contribution of \textit{NN} P-wave becomes almost as important as
one of S-waves in negative parity states for n+$^3$H resonance
region.

In fact low energy nuclear physics is dominated by \textit{NN}
S-waves, being a result of large \textit{NN} scattering lengths
and the fact that interaction in higher waves is weaker.  On
contrary rare physical observables, where \textit{NN} P-waves are
important, have tendency to disagree with the experimental data.
One such example could be the \textit{3N} analyzing powers
\cite{P_waves_Gloe,P_waves_Pisa}. Higher \textit{NN} waves should
contribute in asymmetric nuclear systems; n+$^3$H system due to
its large neutron excess is one such case. Other example can be
given by plotting relative discrepancy in predicted binding
energies by Argonne collaboration \cite{PPWC_PRC64_01} of various
He isotopes (see Fig. \ref{Fig_He_Exp_V18}). The discrepancy
increases linearly with neutron excess, even if \textit{3NF}
improves overall agreement it does not remove this tendency.

All these statements speaks against good description of
\textit{NN} interaction in P-waves. One should also recall that
\textit{NN} interaction is basically tuned on n-p and p-p data.
Moreover, low energy p-p P-waves are overcasted by Coulomb
repulsion, while n-n P-waves are not directly controlled by
experiment at all. We have evaluated, see Fig.
\ref{Fig_He_Exp_V18}, how much n-n P-wave potential must be
changed in order to describe resonance cross sections for Av18
model. It turns to be of order $\sim$20\%, which is rather much to
blame on charge dependence in P-waves as well as compared to
$\sim$6-8\% necessary to solve A$_y$ puzzle in n-d scattering
\cite{P_waves_Gloe,P_waves_Pisa}. However in n-d calculations n-n
and n-p P-waves were modified simultaneously, whereas we have
tuned only n-n P-waves. Still this study suggests that charge
dependence effects can be sizeable in n-n P-waves and can provide
a possible explanation for the disagreement observed in n-$^3$H
resonance region.

\begin{theacknowledgments}
Numerical calculations were performed at Institut du
D\'eveloppement et des Ressources en Informatique Scientifique
(IDRIS) from  CNRS and at Centre de Calcul Recherche et
Technologie (CCRT) from CEA Bruy\`eres le Ch\^atel. We are
grateful to the staff members of these two organizations for their
kind hospitality and useful advices.
\end{theacknowledgments}


\bibliographystyle{aipproc}
\bibliography{sample}

\end{document}